\documentclass[%
 reprint,
showpacs, showkeys, 
 amsmath,amssymb,
 aps,
floatfix,
]{revtex4-1}

\usepackage{graphicx}
\usepackage{dcolumn}
\usepackage{bm}
\usepackage{epstopdf}
\usepackage{color,soul}
\usepackage{setspace}

\usepackage[mathlines]{lineno}

\begin{document}

\preprint{APS/123-QED}

\title{\huge Thermoelectric Energy Harvesting Via Piezoelectric Material} 

\hspace{20 mm} \\
 
\author{Lijie Li}%
 \email{l.li@swansea.ac.uk}
\affiliation{%
 College of Engineering, Swansea University, Swansea, SA2 8PP, UK\\
\\ 
\hspace{1 mm} \\
}%

\date{18 June 2015}

\begin{abstract}
{\large Thermoelectric energy harvesters can have a much higher conversion efficiency by implementing quantum dots/wells between the high temperature region and the low temperature region. However they still suffer a limitation of the maximum output power, represented by the maximum $\Delta E$ (maximum energy gap of two quantum dots/wells layers). In this work, we use the piezoelectric material in the high temperature region, which has conceptually addressed the problem of the maximum power limitation. Full analysis of device physics including comparison with the existing technology and quantum simulation has been conducted to validate this concept. Results show that with the new concept, the maximum output power has been increased by at least an order of magnitude with the same power input and identical device dimensions.}\\

\end{abstract}
{\Large
\pacs{84.60.Rb, 81.07.Ta, 81.05.-t}

\keywords{Thermoelectric energy harvesting, Quantum dots, Piezoelectric}}
\maketitle

\newpage

\section{\large Introduction}
{\large Thermoelectric energy harvesting has the reverse effect as the thermoelectric refrigerator \cite{APL.quantumdots.refrigerator}, and has been studied extensively in recent decades \cite{NM.material} \cite{Science.thermoelectric.cooling.power.generatioon} \cite{Science.thermoelectric} \cite{PNAS.best.thermoelectric}, alongside other notable nanometre sized energy harvesters \cite{ZhonglinWang.Nature} \cite{Qinyong.nature.nano} \cite{Wang201213}. With regards to the recent development in thermoelectric harvesting, several literatures reported that implementing quantum dots or wells into the device can significantly increase device efficiency \cite{PRB.quantumdots} \cite{Nanotech.thermoelectric.quantumdots} \cite{PhysRevLett.most.efficient}, attributed to the energy filtering effect of the quantum dots/wells, which coincides with what has been suggested in \cite{PNAS.best.thermoelectric}. Theoretical investigation of quantum dots based thermoelectric harvesters has been conducted, which was reported in Ref.~\cite{PRB.quantumdots} that the maximum scaled output power with other parameters being optimized appears at around $\Delta E$=6$k_{B}T$, $\Delta E$ being the difference of the energy levels of two quantum dots layers on the left and right sides of the central cavity. Question is then whether there is any solution to overcome this limit. Right from the invention of the seebeck and Peltier effects up to now, people have been using doped semiconductor materials as the central cavity with the aim of increasing the electrical conductivity and reducing thermal conductivity for a higher figure of merit. Specifically for the quantum dots thermoelectric harvesting, very large energy level difference (larger than the optimized $\Delta E$) of quantum dots in left and right sides will cause electrons difficult to enter/exit the central cavity (schematic illustration in Fig.~\ref{fig1}). It is seen from the schematic diagram, the chemical potential of the central cavity generally deployed has been a flat line. Changing the flat potential of the central cavity to a tilted shape will be a conceptual advancement, as ideally the left end of the chemical potential of the central cavity should be align with the energy level of the left quantum dot, and the right end should be align with the right side quantum dot in order to eliminate reflections at the staircase in the potential diagram. It is worth noted that the new concept applies to increasing the efficiency of the thermoelectric cooling as well. Out of many semiconductor materials, a poled piezoelectric material has a ramped chemical potential, which perfectly meets this application. In this article, a piezoelectric material is used for the central hot region. Theoretical investigation conducted here shows that a significant increase in energy efficiency has been observed. Detailed analysis is described and presented in below sections.}

\section{\large Quantum dots thermoelectric harvester}

\begin{figure*}
\includegraphics[width=17cm]{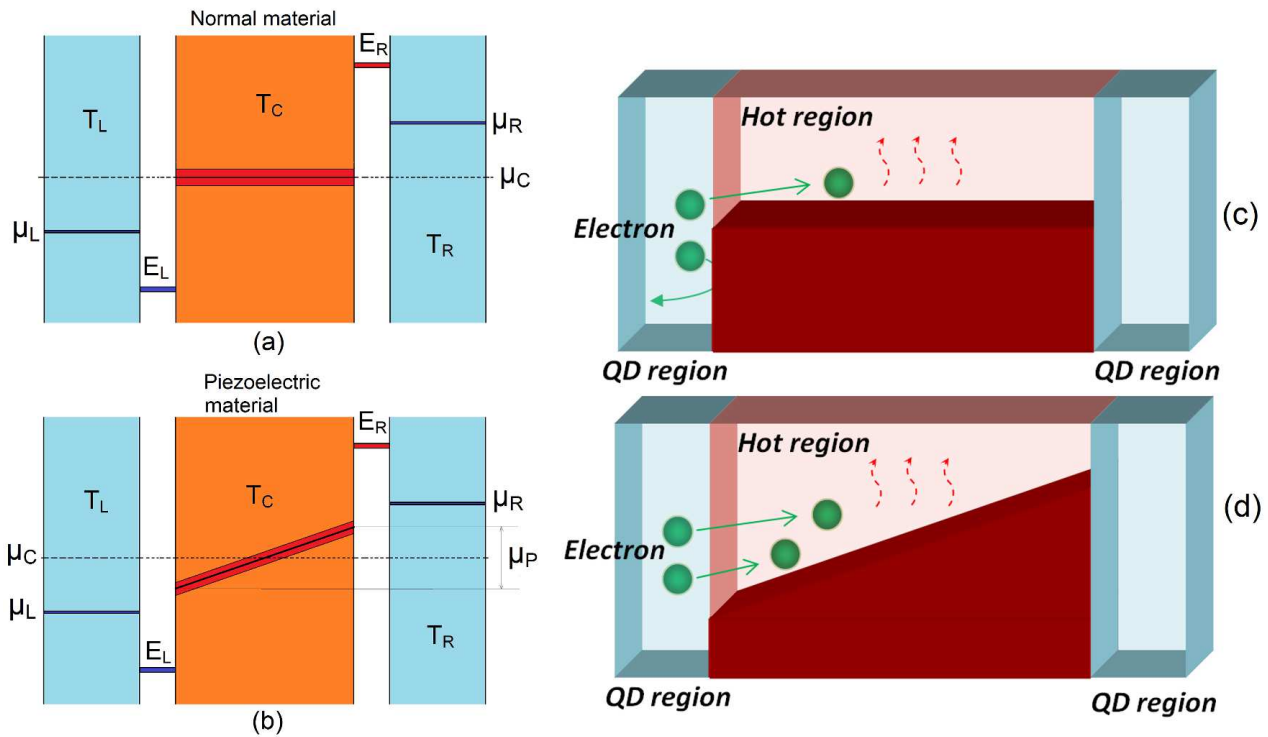}
\caption{ (a) Schematic illustration of the previously reported energy harvester based on quantum dots. Central region is connected to a heat source, where energy of electron has been increased through phonon-electron interaction. (b) Proposed new piezoelectric thermoelectric harvester. Poled piezoelectric material has been used in the central region, so that a chemical potential slope $\mu_{P}$ is present. The chemical potential slope of the piezoelectric layer has significant impact on the maximum output power analysed in the text. (c) 3D illustration of (a) in which electron having lower energy than the central energy level is bounced back. (d) Diagram of the new concept where much more electrons with lower energies can proceed to the central cavity, leading to much improved energy output.}
\label{fig1}
\end{figure*}

\begin{figure}
\includegraphics[width=8cm]{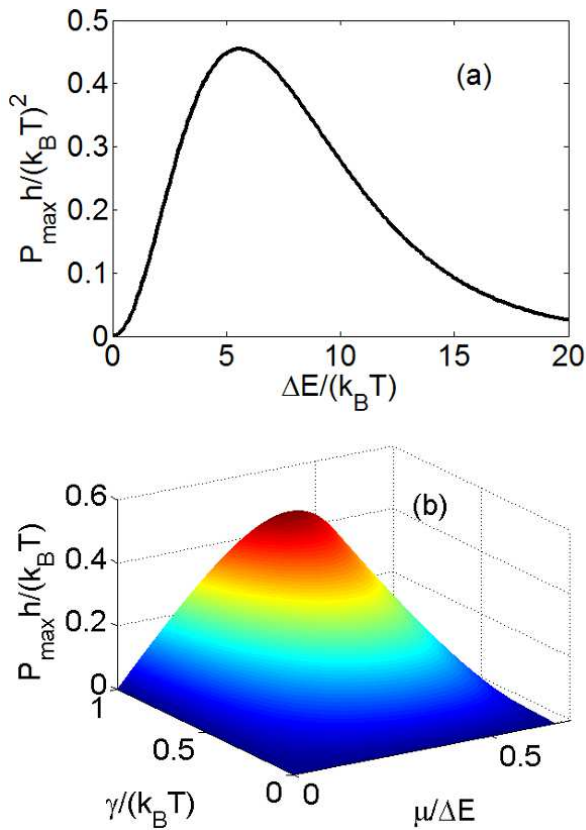}
\caption{ (a) $P_{max}h/(k_{B}T)^2$ versus $\Delta E/(k_{B}T)$ for $\Delta T$ = $T$ and $\mu$ is optimized, and the level width $\gamma$ is chosen as $k_{B}T$. $\Delta T=T_{C}-T_{R}$ and $T=(T_{C}+T_{R})/2$. $\mu_{R,L}=\pm \mu/2+(E_{L}+E_{R})/2$, $\mu_{C}=(E_{L}+E_{R})/2$. $\mu_{C}$ is the chemical potential of the central region. (b) $\Delta E $ is fixed at the optimized value 5.6$\Delta E$, scaled maximum power $P_{max}h/(k_{B}T)^2$ as functions of $\mu$ and $\gamma$.}
\label{fig2}
\end{figure}

{\large In order to understand the motivation of this concept, the optimization of the previously reported quantum dots thermoelectric harvester is demonstrated in this section. As schematically shown in Figs.~\ref{fig1}(a) and 1(c), electrons entering to the central hot area from left side electrons bank gain energy from phonon-electron interaction, and energized electrons flow to the right side of the electrons bank forming a electrical current loop, which operates as a battery. Fig.~\ref{fig1}(c) schematically displays that electrons having lower energies than the central potential will be precluded from entering the central cavity, which is thought the underlying reason of the device being limited by the maximum energy gap $\Delta E$.  Following the procedure in \cite{PRB.quantumdots}, the output power of the harvester can be derived from the conservation laws for electrical charge and energy} 
\begin{align}
&f_{L}-f_{CL}+f_{R}-f_{CR}=0 \nonumber \\
&\frac{Jh}{2\gamma}+E_{L}(f_{L}-f_{CL})+E_{R}(f_{R}-f_{CR})=0
\label{eq:1}
\end{align}\noindent
{\large The first part of the Eq.~(\ref{eq:1}) represents that the total electrical charge equals to zero, which indicates that the charges flowing from the left electrode reservoir to the central cavity equals to the charges flowing from the central cavity to the left side reservoir. Where $f_{L}, f_{CL}, f_{R}, f_{CR}$ are Fermi Dirac distributions characterizing the cavity's occupation function, which are $f_{CL}=f(E_{L}-\mu_{C}, T_{C})$, $f_{L}=f(E_{L}-\mu_{L}, T_{L})$, $f_{CR}=f(E_{R}-\mu_{C}, T_{C})$, $f_{R}=f(E_{R}-\mu_{R}, T_{R})$. The above expressions represent the occupation density of electrons within the energy band (bandwidth is $\gamma$) centred about each energy level. The general form of the Fermi Dirac distribution is $f(E-\mu, T)=1/(1+e^{(E-\mu)/(k_BT)})$. The second part of the Eq.~(\ref{eq:1}) describes that the total energy (external heat energy, heat current from the left to the right reservoirs) is zero. $h$ is the Planck's constant. Re-organize the Eq.~(\ref{eq:1}), one can get the heat current $J$}
\begin{equation}
J=\frac{2\gamma \Delta E}{h}(f_{CR}-f_{R})
\label{eq:2}
\end{equation} \noindent
The electrical current $I$ is 
\begin{equation}
I=\frac{eJ}{\Delta E}=\frac{2e\gamma (f_{CR}-f_{R})}{h}
\label{eq:3}
\end{equation}\noindent
{\large where $e$ is the charge of an electron, and $\Delta E = E_{R}-E_{L}$. The efficiency of the system defined as the ratio of the harvested power $P=|(\mu_{L}-\mu_{R})I|/e$ to the heat current $J$. $\mu_{L,R}$ are chemical potentials of the left and right electrodes. The efficiency $\eta$ is then expressed as $\eta=(|\mu_{L}-\mu_{R}|)/ \Delta E$. It is then derived that the $\mu=|\mu_{L}-\mu_{R}|$ $\leq$ $\Delta E$, as the $\eta \leq 1$. The output power is expressed as}

\begin{equation}
P=\frac{2\mu \gamma}{h}(f_{CR}-f_{R})
\label{eq:4}
\end{equation}\noindent
{\large Numerical simulation can be performed to arrive at the maximum output power as the function of the $\Delta E$. In the simulation, $\mu$ and $\gamma$ have been optimised to have the maximum $P$. Simulated results are shown in Fig.~\ref{fig2}. For the parameters set in the figure caption, the maximum $P$ is at around 6$k_{B}T$, at which the optimized $\mu$ is around 0.43$\Delta E$, and the $\gamma$ = $k_{B}T$. The above results match closely with what was reported in \cite{PRB.quantumdots}. With the question of increasing the maximum power, it is envisaged that a ramped chemical potential may result in a better performance, as electrons with lower energy can have the opportunity to enter the central region (Fig.~\ref{fig1}(d)). Device physics of the new configuration is described in the following section.}

\section{\large Device Concept Incorporating Piezoelectric Material}
\begin{figure}
\includegraphics[width=8cm]{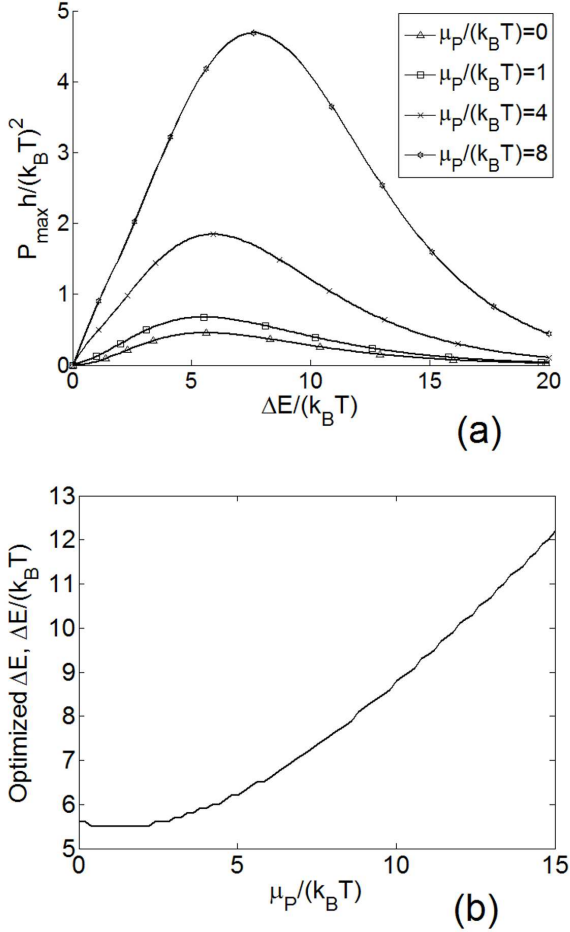}
\caption{ (a) Calculated maximum output power $P_{max} h/(k_{B}T)^2$ relating to energy gap between two layers of quantum dots $\Delta E/(k_{B}T)$ for various fixed piezopotentials $\mu_{P}/(k_{B}T)$. (b) Optimized $\Delta E$ in relation to the piezopotentials $\mu_{P}$.}
\label{fig3}
\end{figure}

\begin{figure}
\includegraphics[width=8cm]{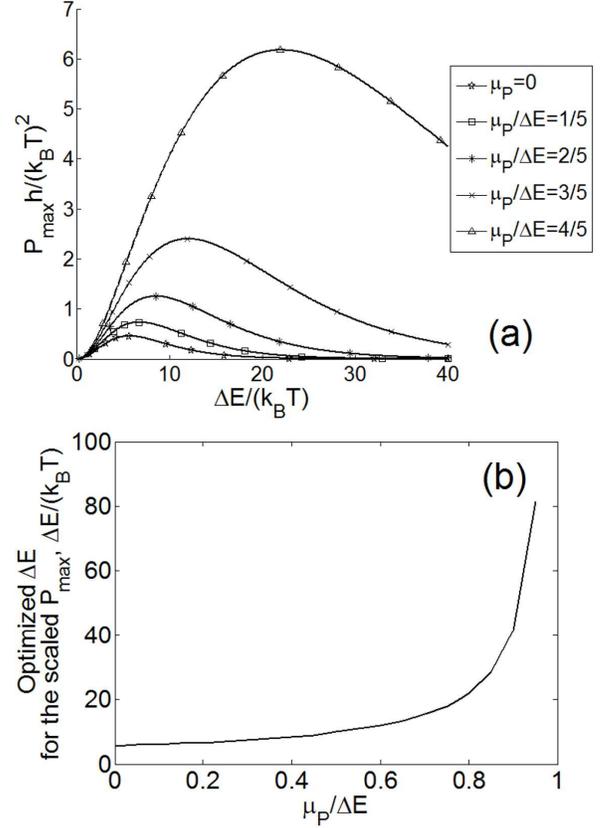}
\caption{ (a) Maximum output power versus energy band gap of quantum dots for various piezopotentials proportional to $\Delta E$. (b) Optimized $\Delta E$ versus the ratio of $\mu_{P}$ to $\Delta E$.}
\label{fig4}
\end{figure}

{\large In the new device architecture (Figs.~\ref{fig1}(b) and 1(d)), with the aim of overcoming the limitation of the maximum output power obtained from the previous analysis ($\sim$6$k_{B}T$), a piezoelectric material has been chosen to make the central cavity. For a poled piezoelectric material, the chemical potential is no longer a flat line, instead a slope \cite{SR.piezoelectric.potential}, whose gradient is dependent on the material properties. The potential gradient of the poled piezoelectric materials is understood from aligned electrical dipoles. A piezoelectric material can have both electrical conductivity and piezoelectricity simultaneously, for example bismuth titanate (BIT) \cite{JACS}. Moreover due to the high transition
temperature, BIT ceramics are good candidates for
high temperature piezoelectric applications \cite{JECS}. Well processed piezoelectric material will meet the requirement for the hot cavity in the thermoelectric harvester in the viewpoint of practical realization. The energy levels of the quantum dots can be tuned by the external bias, alternatively they can be re-configured by the externally applied mechanical stress \cite{JAP.strain.induced.quantum.dots} \cite{APL.MEMS.modified.quantum.dots}. In the new device, the chemical potential of the central region becomes $\mu_{C}^{U,L}=(E_{L}+E_{R})/2\pm\mu_{P}/2$, where the $\mu_{C}^{U,L}$ are the upper and lower ends of the chemical potential of the piezoelectric layer. $\mu_{P}$ is the difference between the upper and lower ends of the chemical potential of the piezoelectric material. Hence the $f_{CR}$ becomes}

\begin{equation}
f_{CR}=\frac{1}{\exp[(\Delta E-\mu_{P})/(3k_{B}T)]}
\label{eq:5}
\end{equation} \noindent
{\large Substituting the Eq.~\ref{eq:5} to the Eq.~\ref{eq:4}, numerical analysis has been conducted using optimized $\mu$ and $\gamma$. The $\mu_{P}$ independent of $\Delta E$ has been initially designated to fixed values as the function of $k_{B}T$. The results (shown in Fig.~\ref{fig3}) display a significant increase of the scaled maximum power $P_{max}h/(k_{B}T)^2$ from around 0.4 to 5 with the $\mu_{P}$ increasing from 0 to 8$k_{B}T$. However it does not demonstrate the similar scale of the increase for the optimized $\Delta E$, which can be observed to increase from around 5.6$k_{B}T$ to around 12$k_{B}T$ as $\mu_{P}$ increasing from 0 to 15$k_{B}T$. This is due to that the $\mu_{P}$ is given fixed values while the $\Delta E$ varies. Further investigation is therefore performed with $\mu_{P}$ being dependent upon the $\Delta E$. In the analysis, the $\mu_{P}$ has been designated as a ratio of $\Delta E$, i.e. from 1/5 of $\Delta E$ to 4/5 of the $\Delta E$. Numerical calculation (shown in Fig.~\ref{fig4}) demonstrates that apart from the obvious increase of the maximum power, the optimized $\Delta E$ has also increased to around 80$k_{B}T$ for $\mu_{P}=0.9\Delta E$. Extended analysis has been conducted, which shows that when the $\mu_{P}$ approaches to $\Delta E$, there is no maximum $\Delta E$, instead the maximum power continues increasing with the increasing $\Delta E$.}

\begin{figure}
\includegraphics[width=9cm]{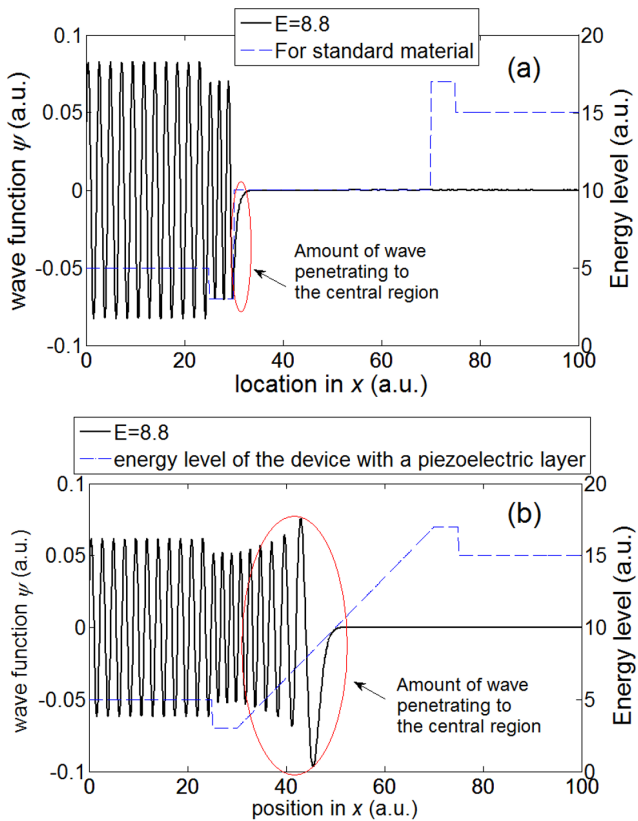}
\caption{ Time-independent analysis. To simplify the calculation, let $\hbar$=1, $m$=1, length in $x$ direction is 100. $\Delta E$=14, $\mu_{P}=\Delta E$, $\mu$=10. (a) $\psi$ versus position in $x$ and the energy diagram of device with a standard material. (b) $\psi$ versus position in $x$ and the energy diagram of device with a piezoelectric material.}
\label{fig5}
\end{figure}

\begin{figure*}
\includegraphics[width=17cm]{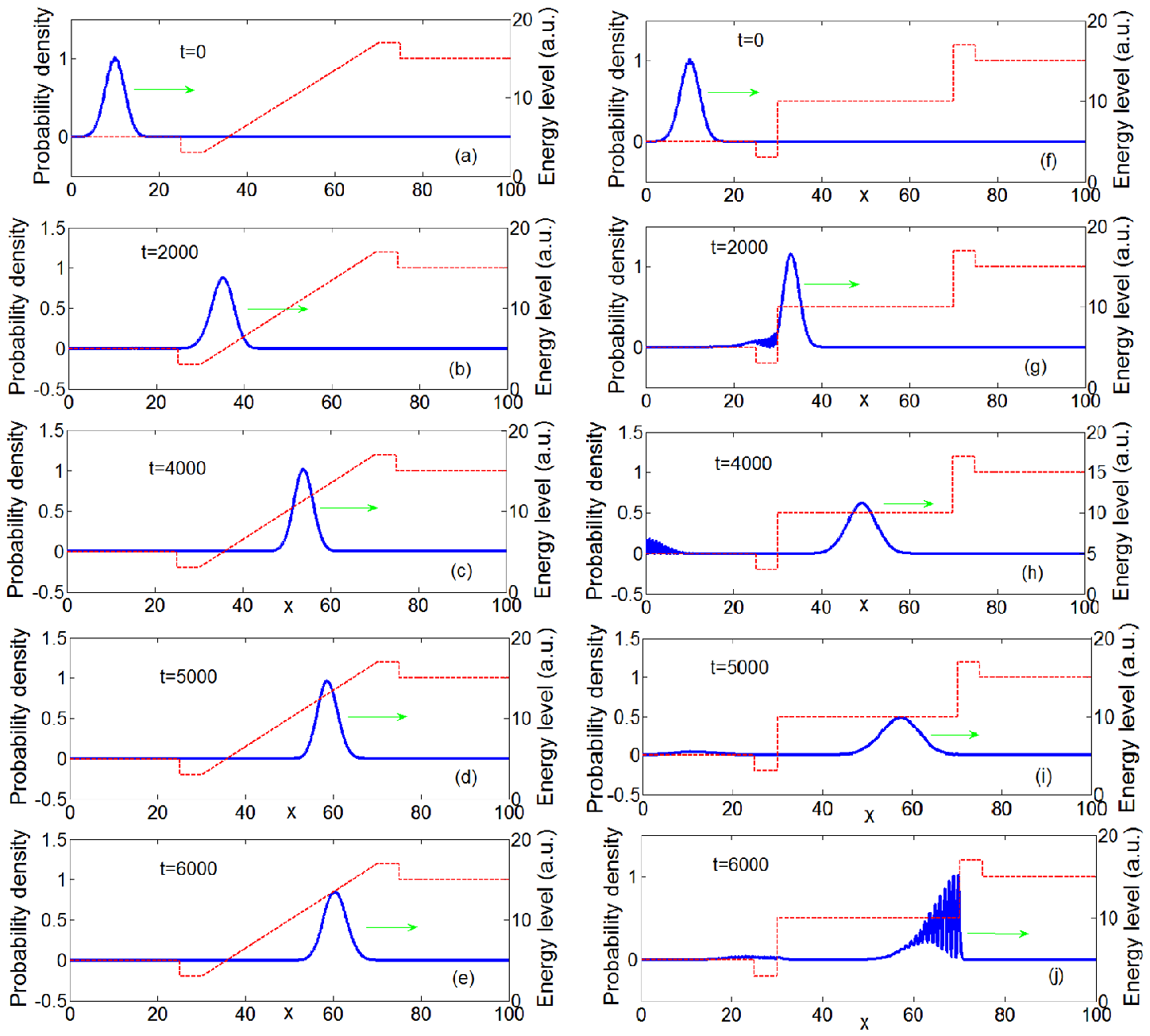}
\caption{ Time-dependent quantum analysis. To simplify the calculation, let $\hbar$=1, $m$=1, length in $x$ direction is 100. $\Delta E$=14, $\mu_{P}=\Delta E$, $\mu$=10. (a)-(e) moving wave packet from the left to the right, energy level of electron is 8.8 a.u., at time points of 0, 2000, 4000, 5000, and 6000 respectively. This is for the device with a piezoelectric material. (f)-(j) moving wave packet of the electron (E=8.8 a.u.) for the device with normal material in the central. Starting position of both cases is from x=10, initial amplitude of both the wave packets is set to 1, initial width of the wave packet is 20, $\Delta x$=0.1, and $\Delta t$=0.003.}
\label{fig6}
\end{figure*}

\section{\large Quantum Analysis}

{\large Further analysis is needed using quantum mechanics theory to reveal the physics behind the performances of the new proposed device. As the fundamental element of the device is a quantum dot, quantum theory has to be used \cite{LI}. The core of the quantum theory is based on the Schrodinger equation, here both the time-dependent and time-independent Schrodinger equations are used in the simulation, which are shown in below}

\begin{equation}
i\hbar\frac{\partial}{\partial t}\psi(x,t)=[-\frac{\hbar^2}{2m}\nabla^2+V(x,t)]\psi(x,t)
\label{eq:6}
\end{equation}\noindent
{\large where $\psi (x,t)$ is the wave function of an electron as the function of time $t$ and location $x$, $\hbar$ is Planck's constant divided by 2$\pi$. $m$ is the mass of the electron and $V(x,t)$ is the potential diagram through which the electron passes. The time-independent Schrodinger equation simplified from the Eq.~\ref{eq:6} is}
\begin{equation}
E\psi(x)=[-\frac{\hbar^2}{2m}\nabla^2+V(x)]\psi(x)
\label{eq:7}
\end{equation}\noindent
{\large where only the static state of the probability along the location x of an electron with certain energy level $E$ can be calculated. One dimensional numerical simulation has been conducted using Eqs.~\ref{eq:6} and \ref{eq:7}. Results from calculations have been shown in Figs.~\ref{fig5} and \ref{fig6}. Fig.~\ref{fig5} shows the calculated distribution of the wave function $\psi$ for the electron with the initial energy level of 8.8 a.u. along both the energy diagrams representing the device with a traditional material and the device with a piezoelectric material respectively. All parameters used in the simulation are defined in the figure caption of the Fig.~\ref{fig5} It can be summarized that for the standard material there is very little wave penetrating to the central region as the energy level of the central material is 10, which is higher than the electron energy level. However for the device constructed by a piezoelectric material, there is a good portion of wave passing to the central region as highlighted by the red elliptic shape in Fig.~\ref{fig5}(b). It should be noted that the amount of wave through to the central region will be energized by the external heating source and eventually going to the left side of the device forming an electrical current loop with external loads. While for the device with the normal material, electrons with less energy than the central energy level will die out immediately. Fig.~\ref{fig6} shows the dynamic simulation as to an electron passing through the two potential diagrams using the one dimensional time-dependent Schrodinger equation. At a sequence of time points, the wave packet travels along the axis $x$ for the two different energy diagrams have been calculated using simplified parameters. At the beginning, two cases are pretty much similar. A little portion of the wave packet has been reflected at the potential step of the device with the central cavity made of a traditional material. In comparison, it is a very smooth transition without any reflection on the potential diagram of the piezoelectric material. As the time develops further it can be seen that in the normal material, the wave packet travels faster than in the piezoelectric material, leading to less phonon-electron interaction time.  
It is seen from the results that three advantages are demonstrated by the novel piezoelectric quantum dots thermoelectric energy harvester: 1), more electrons with lower energy can reside in the central region, evidenced by the time-independent Schrodinger analysis. 2), less reflection can be achieved, supported by the time-dependent Schrodinger analysis. 3), slower wave packet travelling speed offering longer thermoelectric transferring time is observed.
It was described in the \cite{PRB.quantumdots} that with the accumulative effect taken into account, the quantum dots thermoelectric energy harvester could reach the value of 0.1W per cm$^2$ at temperature difference of 1K. From above analysis , the new piezoelectric material based quantum dots thermoelectric harvester could have the maximum power output of more than 1W per cm$^2$, which presents the highest efficiency for the energy harvesters of this kind.}

\section{\large Conclusion}
{\large To conclude, with the objective of achieving higher power efficiency of the quantum dots thermoelectric energy harvester, a piezoelectric material is used to replace the standard material in the high temperature region. According the analysis, the power output of the new device will be at least 10 times higher. Quantum mechanical analysis reveals that with a tilted chemical potential of the piezoelectric layer, the electron passes through the central region with minimal reflection, and more electrons are energized by the thermal energy in the central hot region.}

\bibliography{myrefs}

\end{document}